# PREVENTION IS BETTER THAN CURE: EXPERIMENTAL EVIDENCE FROM MILK FEVER INCIDENCE IN DAIRY ANIMALS OF HARYANA, INDIA

A. G. Adeeth Cariappa*[1], B. S. Chandel[1], Gopal Sankhala[2], Veena Mani[3], Sendhil R[4], Anil Kumar Dixit[1] and B. S. Meena[2]

## ABSTRACT

Calcium deficiency in high yielding bovines during calving causes "milk fever" which leads to economic losses of around ₹ 1,000 crores (US $ 137 million) per annum in Haryana, India. With increasing milk production, the risk of milk fever is continuously rising. In the context, we aim to address the most fundamental research question: What is the effect of a preventive health product (anionic mineral mixture (AMM)) on milk fever incidence, milk productivity and farmers' income? In an effort to contribute to the scanty economic literature on effect of preventive measures on nutritional deficiency disorders in dairy animals, specifically, on AMM effects in India, this study uses a randomized controlled design to estimate internally valid estimates. Using data from 200 dairy farms, results indicate that milk fever incidence decreases from 21% at baseline to 2% in treated animals at follow-up. Further, AMM leads to a 12% and 38% increase in milk yield and farmer's net income, respectively. Profits earned due to the prevention of milk fever [₹ 16,000 (US$ 218.7)] overweighs the losses from milk fever [₹ 4,000 (US$ 54.7)]; thus, prevention using AMM is better than cure.

**KEYWORDS** Milk fever; anionic mineral mixture; randomized controlled trial; impact evaluation; farmers' income; dairy

**JEL codes** Q10, Q120, Q180, C93

---

[1] Division of Dairy Economics, Statistics and Management, ICAR–National Dairy Research Institute (ICAR-NDRI), Karnal – 132001 (Haryana), India.
* Correspondence: adeeth07@gmail.com
[2] Division of Dairy Extension, ICAR–NDRI – 132001 (Haryana), India
[3] Division of Animal Nutrition, ICAR–NDRI – 132001 (Haryana), India
[4] ICAR-Indian Institute of Wheat and Barley Research, Karnal - 132 001 (Haryana), India

**INTRODUCTION**

Nutritional deficiency disorders in dairy animals and the damages associated with it are avoidable depletion of scarce resources (Thirunavukkarasu et al. 2010). Economic implications of these have assumed increased significance with rapid commercialization and intensification of dairy farming in India. Nutritional deficiency disorders (like calcium or magnesium deficiency) affect the productive and reproductive performance of dairy animals causing economic losses to farmers. Increased loss of milk due to nutritional disorders leads to decreased availability and increased purchase costs to consumers threatening nutritional security of the producers (Hogeveen, Steeneveld and Wolf 2019; Jodlowski et al. 2016; Nilsson et al. 2019). Economic literature provides evidence on the spread and persistence of bacterial and viral dairy animal diseases (Hayer et al. 2017; Sok and Fischer 2020), losses associated with it (Gohin and Rault 2013; Govindaraj et al. 2017; Govindaraj et al. 2021; Boisvert, Kay and Turvey 2012; Barratt et al. 2018) and the control/prevention strategies (Chi et al. 2002; Rich and Winter-Nelson 2007; Hennessy 2007; Rich, Roland-holst and Otte 2014; Fadiga and Katjiuongua 2014; Wang and Hennessy 2015; Schroeder et al. 2015; Krieger et al. 2017; Wang and Hennessy 2014). However, evidence on the economic effects of prevention of nutritional deficiency disorders is absent.

Milk fever (MF), hypocalcemia, is one such metabolic illness in dairy animals. It is the decrease in blood calcium (Ca) level due to rapid drain of Ca into colostrum after parturition (calving) (Melendez and Poock 2017; Rodríguez, Arís and Bach 2017). As Ca is essential for organs which has smooth muscle functions like uterus, abomasum, teat sphincter and rumen; the deficiency of Ca or MF affects these organs hampering fertility and productive performance (Melendez et al. 2019; Caixeta et al. 2017). MF also leads to immune suppression which makes animals susceptible to economically important diseases like mastitis (Kimura, Reinhardt and Goff 2006). The risk of other metabolic disorders like dystocia, uterine prolapse, retained fetal membranes, ketosis, metritis, *etc* increases with the presence of MF (Goff 2008; Reinhardt et al. 2011; Melendez and Poock 2017; Oetzel and Miller 2012).

MF is economically important as the incidence, and thus losses, increases with age; in the USA, MF incidence ranges from 25% to 42% in $1^{st}$ to $6^{th}$ lactation animals (Reinhardt et al. 2011). In India, data on MF incidence is scanty and poorly recorded at the national level. However, three surveys documented the incidence to be 11-12% in north-eastern states (Paul, Chandel and Ray 2013), 13-14% in Tamil Nadu (Thirunavukkarasu et al. 2010) and 10% in Himachal Pradesh (Thakur et al. 2017). The losses due to MF in Tamil Nadu was estimated at around ₹ 41 crores (US$ 5.4 million) during 2005-08 (Thirunavukkarasu et al. 2010). It was estimated it to be around US$ 137 million (in high risk animals) during 2020 in Haryana (Cariappa et al. 2021).

MF prevention is therefore indispensible for the success of reproductive and productive performance of dairy animals Melendez and Poock (2017) and also from the economic point of view (Cariappa et al. 2021). We can avoid loss in milk production and enhance farmers' income by preventing MF, which ensures food and nutritional security of the dairy farmers. It assumes paramount importance now as nutritional outcomes are deteriorating evident in the fall of India's rank in Global Hunger Index[5].

Although metabolic disorders are a result of complex interactions between risk factors, it can be prevented by the right management decisions (Krieger et al. 2017). Blanc et al. (2014) discusses various approaches available for preventing the MF; the first is feeding anionic salts (negative dietary cation-anion difference) before parturition and second, oral or intravenous Ca supplementation instantly after calving. Benefits of the latter require further investigation as the evidence supporting it is not conclusive (Blanc et al. 2014). The dietary cation-anion difference (DCAD) of dairy animals was originally manipulated to combat the MF (Ender, Dishington and Helgebostad 1971; Block 1984; Delaquis and Block 1995; Iwaniuk and Erdman 2015). Several meta-analysis and systemic reviews of experiments conclude that feeding anionic feed (negative DCAD) pre-partum improves Ca concentration and reduces MF while improving reproductive and productive performance of dairy animals (Oetzel 1991; Lean et al. 2006; Charbonneau,

---

[5] https://www.globalhungerindex.org/india.html

Pellerin and Oetzel 2006; Santos et al. 2019; Lean et al. 2019). A very few studies beg to differ from these conclusions (Ramos-Nieves et al. 2009).

In 2016, the Indian Council of Agricultural Research–National Dairy Research Institute (ICAR–NDRI), developed and commercialized an 'anionic mineral mixture' to prevent MF in dairy animals[6]. Till date, except an observational study by Thakur et al. (2017), there is no study that evaluates the effect of anionic diets on animal health or production parameters outside the controlled setting of experimental farms of Indian research institutes. Thakur et al. (2017) observed that the MF incidence decreases from 10% in non-users to 2.5% in users of anionic diets; milk yield improvement was observed in 60% (in 24 out of 40 animals) and 45% of the animals in user and non-user group respectively. However, the study did not quantify the level of yield improvement.

In this study, we go beyond primary trials in research institutes and observational studies. Our work is set in five villages of Haryana State, Northern India with 200 animals from 200 dairy farms (100 cows and 100 buffaloes). In an effort to produce internally valid estimates, we use a randomized controlled design to conduct a household level analysis of the effect of feeding anionic diets to prevent MF in farmers' field. We complement the vast literature to include the economic effects on self-reported incidence of MF, milk yield, fat-corrected milk yield and farmer's income. We show that AMM is an effective strategy in preventing MF and increasing milk productivity and farmers' net return.

---

[6] https://kamdhenufeeds.com/product/anionic-mishran/
http://ndri.res.in/technologies-patents/

The following section describes the intervention followed by study and intervention design. The next section outlines our data and sample characteristics followed by balance test results. The estimation strategy is detailed after that. Finally, we report and discuss our impact estimates before concluding the study.

**INTERVENTION**

The 'Anionic Mineral Mixture' (AMM) is designed to reduce MF and other post-partum problems in cows and buffaloes. The technology along with anions contains Vitamin E, useful against oxidative stress in the pregnant cows which makes it resistant to metabolic disorders and increases reproductive performance[7]. The supplementation of AMM is said to benefit the dairy farmers economically by improving the milk yield (by 10%), fat content of the milk and the immunity of the animals apart from preventing various diseases[8]. The AMM contains 7640 mEq and 5080 mEq anionic value of sulphur and chloride, respectively; 1340 mEq cationic value of potassium with a total negative dietary cationic anionic difference (DCAD) of 11,380 mEq and 10,000 IU/kg of Vitamin E. 50 grams each in the morning

---

[7] See Appendix 1 on how AMM causes decrease in MF incidence and increase in productive and reproductive performance of dairy animals.
[8] https://kamdhenufeeds.com/product/anionic-mishran/

and evening before 3-4 weeks of calving is the recommended dosage of AMM feeding.

*Study design*

We begin with the sample size determination using statistical power calculations. This is based on one of the primary outcomes – milk yield per animal per day – as registered in the pre-analysis plan[9]. Due to unavailability of data on the other primary outcome – incidence of MF – power calculations were done only based on the milk yield. These calculations are designed to give an 80% chance to correctly detect the effect when there is an effect (power), with 5% level of significance. Data on mean of milk yield in rural Haryana is obtained from Indiastat database[10] and standard deviation is taken from a survey conducted by (Lal et al. 2020). Using the above parameters and an assumed $R^2$ of 0.5 in the final impact regression[11], a minimum total sample size of 172 animals was required to detect a statistically significant effect of 10% increase in milk yield between treated and control groups. To account for the possibility of attrition, a sample of 200 animals (100 cows and 100 buffaloes) with 100 in treatment (50 cows and 50 buffaloes) and 100 in control (50 cows and 50 buffaloes) were included. Additional 14 animals were sampled, in case of replacement. Note that 200 eligible animals were selected from 200 different dairy farms.

---

[9] "Impact evaluation of anionic mineral mixture supplementation on milk production and the milk fever: a randomized control trial." AEA RCT Registry. April 17. https://doi.org/10.1257/rct.5108-2.0
[10] https://www.indiastat.com/data/agriculture/milk-production/data-year/2019
[11] The $R^2$ in final regression was around 0.7.

The AMM is aimed at reducing the MF in high yielding dairy animals. For this reason, we needed to work in areas where milk yield is high and the population of high yielding animals is large. The funding agency of the study has adopted 5 villages namely Samora, Garhi Gujran, Churni, Kamalpur Roran and Nagla Roran in the Karnal district of Haryana state in India. As Karnal is home for around 2.8 lakh high yielding (1.1 lakh exotic/crossbred cows and 1.8 lakh buffaloes) female bovines (Government of Haryana 2020), these villages were ideal for our experiment[12]. Data was collected from farmers who owned animals with high risk of MF incidence (animals in second or above parity with peak milk yield of more than 10 kg/day), which are not fed any type of anionic diets and had at least 1 month before parturition (calving)[13]. Baseline survey was conducted during September 2020 and the intervention (AMM) was supplemented to the treatment group during September-November 2020. The follow-up survey was done 2-3 months post-parturition between the last week of January and first week of February 2021.

*Intervention design*

On the demand side, a majority (77%) of the farmers were aware of MF but weren't aware of the AMM. Only around 3% (7 out of 214) of the farmers knew about AMM and only 1 farmer had used it earlier. Around half of the respondents report of taking precautionary measures against MF like feeding calcium solutions, jaggery or combination of both post-partum (after calving). However, the use of preventive measures pre-partum (before calving) is absent in the sample villages. On the supply side, there is no availability of AMM in the villages or nearby sub-urban centers or in the district headquarters (which is around 20-25 km away from the village). This by default ensured that animals are not fed with any form of anionic diets satisfying the inclusion criterion.

---

[12] The AMM was not supplemented to the animals in these villages; also, both the project staff and villagers were unaware of the product.
[13] Flow chart of the sampling plan is presented in Appendix 2.

The animals, after stratification (cows and buffaloes), were randomly assigned to treatment and control groups using the random number generator in Stata statistical package[14]. The intervention had two components; creating awareness and AMM supplementation. Creating awareness started right away from the baseline survey. All the farmers were given a brochure in local language (i.e., in Hindi) explaining the benefits of AMM, type of animals susceptible to MF and dosage, which they stuck on the walls of their farm. Also, subject matter specialists visited all the villages explaining the importance of MF prevention and how AMM helps to do it. The second and important component is AMM supplementation. AMM was supplemented to all the animals in the treatment group with specific instructions on how to use the product. The control group farmers were promised that they will also get AMM at a later date to avoid resentment against the institute. This is essential because the study is based on individual level randomization; treated and control farmers are neighbors and the control farmers could attrite. Therefore, control farmers were taken into confidence before starting the intervention. After 60 days of parturition of treated animals, follow-up survey was undertaken and soon after, the control animals were also supplemented with AMM. Regular monitoring of treatment was done through field visits and telephonic conversations to confirm that AMM is supplemented to animals properly and in-time.

---

[14] https://dimewiki.worldbank.org/Randomization_in_Stata

## DATA AND DESCRIPTIVE STATISTICS

*Primary outcomes*

As specified in the pre-plan, this study has two primary outcomes; incidence of MF (1/0) and milk yield (kg/animal/day). These are selected based on the goals of AMM – reducing MF and increasing the milk yield.

*Incidence of MF*: Ca deficiency in high yielding bovines upon parturition leads to a condition called MF. It occurs usually within 72 hours of calving. The animals with MF express the following clinical signs. It will be unable to stand up and will be lying down, with its neck turned to one side and then laterally. In severe cases, animal becomes unconscious with sub-normal temperature and if left untreated, it will succumb (NDDB 2019). Farmers were asked if they observed these symptoms both at the baseline and follow-up. For analytical purposes, animals with clinical MF are coded 1 and zero otherwise.

*Milk yield (productivity):* At the baseline and follow-up, we recorded the peak milk yield in the previous lactation. The peak milk yield was converted into average daily milk yield by using the standard conversion factor[15]. The unit of measurement is kg/animal/day.

*Secondary outcome*

*Income from dairying:* We estimate the net income from milk produced. The net income is calculated by subtracting variable costs (sum total of feed and fodder costs, veterinary costs like expenses on artificial insemination, vaccination and deworming, hired labor costs and MF treatment costs) from the value of milk produced (product of price of milk and total lactation milk production[16]). The unit of measurement is ₹/animal/lactation.

*Sample characteristics*

Table 1 presents the description of dependent variables, household characteristics, animal characteristics and feeding regime in the sample. In brief, dairying is the principal source of income for the sample farmers (67%). Treatment and control group farmers own around 6 to 7 acres of land and 5 to 6 dairy animals. Control and treatment group farmers have a median education score of 4 and 5 respectively, implying they have attended school only upto secondary (9-10) and higher secondary (11-12) level.

---

[15] $Average\ daily\ milk\ yield = peak\ yield \times \frac{200}{lactation\ length}$
[16] $Lactation\ milk\ production = average\ daily\ milk\ yield \times lactation\ length$

Average daily milk yield in buffaloes and cows is around 8 and 10-11 kg/animal, respectively at the baseline. The peak milk yield in the present lactation is around 13 and 17 kg/animal. All the animals in the sample are on average in their 3rd parity. The incidence of MF at the baseline is around 15% and 27% in cows and buffaloes, respectively. Net income per annum from dairying (net returns to variable costs) is higher for buffalo rearers (around ₹ 49,740/animal) than cow rearers (₹ 46,030/animal). The average variable cost is a little higher for crossbred cows than buffaloes. The price realized by farmers, as expected, was higher for buffalo milk (₹ 47/kg) than cow milk (₹ 31/kg). Although the prices realized by the treatment group farmers was significantly (slightly) higher (₹ 30.6/kg) than control farmers (₹ 31.3/kg), the value of milk sold was statistically similar. The summary of baseline characteristics reflects the inclusion criterions as we had selected higher parity animals with a peak milk yield of at least 10 kg/day.

The control variables used in our analysis include green fodder fed, dry fodder fed, concentrates fed and herd size. On an average, green fodder, dry fodder and concentrate fed per dairy animals is around 20-21 kg/day, 12 kg/day and 4 kg/day, respectively in the sample area. Control group cows are fed significantly higher green fodder than the treatment group cows.

*Balance*

Although statistical similarities in individual variables are achieved, sometimes the differences in treatment and control group characteristics might be in the same direction. This is an indication of the inability of the random assignment to generate two statistically similar groups. A solution is to complement Table 1 with a test for joint orthogonality (McKenzie 2015). Linear probability estimates of correlates of treatment status shows that the relationship between the treatment status and control as well as dependent variables are non-significant except for net income and variable costs incurred (at 10% level) (Table 2). Joint test of orthogonality (F test) indicates that random assignment to two groups has succeeded in generating balance. Under pure randomization, if balance is achieved in observed variables we can expect to have balance in unmeasured/unobserved variables (Bruhn and McKenzie 2009).

**Table 1 Baseline characteristics of sampling units and mean difference by treatment status**

| Variables | Buffalo (n=100) | | | | | Cow (n=100) | | | | | Overall (n=200) | | | | |
|---|---|---|---|---|---|---|---|---|---|---|---|---|---|---|---|
| | Control (n=50) | | Treated (n=50) | | Mean Diff (C-T) | Control (n=50) | | Treated (n=50) | | Mean Diff (C-T) | Control (n=100) | | Treated (n=100) | | Mean Diff (C-T) |
| | Mean | SD | Mean | SD | | Mean | SD | Mean | SD | | Mean | SD | Mean | SD | |
| **Panel A. Dependent variables** | | | | | | | | | | | | | | | |
| Average milk yield (kg/animal/day) | 7.80 | 1.94 | 7.86 | 2.82 | -0.06 | 10.19 | 3.84 | 10.72 | 4.57 | -0.53 | 8.99 | 2.45 | 9.29 | 4.25 | -0.30 |
| Incidence of MF (1/0) | 0.18 | 0.37 | 0.12 | 0.26 | 0.06 | 0.28 | 0.44 | 0.26 | 0.35 | 0.02 | 0.23 | 0.32 | 0.19 | 0.40 | 0.04 |
| Net income (000' ₹/animal/lactation) | 51.22 | 22.26 | 54.56 | 23.18 | -3.34 | 40.85 | 33.72 | 44.91 | 39.23 | -4.06 | 46.03 | 28.90 | 49.74 | 32.42 | -3.70 |
| **Panel B. Household characteristics** | | | | | | | | | | | | | | | |
| Experience in dairying (years) | 13.22 | 9.55 | 14.20 | 10.96 | -0.98 | 14.76 | 7.92 | 15.26 | 9.61 | -0.50 | 13.99 | 10.26 | 14.73 | 8.79 | -0.74 |
| Education (1-7)[+] | 4.36 | 1.42 | 4.22 | 1.89 | 0.14 | 4.10 | 1.74 | 4.06 | 1.58 | 0.04 | 4.23 | 1.59 | 4.14 | 1.74 | 0.09 |
| Household size (nos.) | 7.33 | 3.76 | 6.34 | 2.83 | 0.99 | 6.10 | 2.74 | 6.84 | 3.35 | -0.73 | 6.70 | 3.32 | 6.59 | 3.10 | 0.11 |
| Training in dairying (1/0) | 0.26 | 0.44 | 0.24 | 0.43 | 0.02 | 0.32 | 0.47 | 0.38 | 0.49 | -0.06 | 0.29 | 0.46 | 0.31 | 0.46 | -0.02 |
| Principal income from dairying (1/0) | 0.66 | 0.48 | 0.52 | 0.50 | 0.14 | 0.68 | 0.47 | 0.82 | 0.38 | -0.14 | 0.67 | 0.47 | 0.67 | 0.47 | 0.00 |
| Land owned (acres) | 6.39 | 6.61 | 6.41 | 5.63 | -0.02 | 5.12 | 4.58 | 6.83 | 5.48 | -1.71* | 5.76 | 6.12 | 6.62 | 5.11 | -0.86 |
| **Panel C. Animal characteristics** | | | | | | | | | | | | | | | |
| Parity (nos.) | 2.64 | 0.80 | 2.80 | 0.81 | -0.16 | 2.80 | 1.03 | 2.80 | 0.90 | 0.00 | 2.72 | 0.81 | 2.80 | 0.97 | -0.08 |
| Peak milk yield in the previous lactation (kg/animal/ day) | 11.89 | 2.95 | 11.98 | 4.30 | -0.09 | 15.54 | 5.86 | 16.35 | 7.02 | -0.81 | 13.72 | 3.74 | 14.17 | 6.51 | -0.45 |
| Peak milk yield in the present lactation (kg/animal/ day) | 12.76 | 2.62 | 12.86 | 3.93 | -0.10 | 16.62 | 6.23 | 17.23 | 6.87 | -0.61 | 14.69 | 3.40 | 15.05 | 6.59 | -0.36 |
| Herd size (nos.) | 6.24 | 3.18 | 6.30 | 2.82 | -0.06 | 4.60 | 2.95 | 5.10 | 2.56 | -0.50 | 5.42 | 3.00 | 5.70 | 2.77 | -0.28 |
| **Panel D. Feed and fodder fed** | | | | | | | | | | | | | | | |
| Green fodder fed (kg/animal/ day) | 21.65 | 6.58 | 20.49 | 6.30 | 1.16 | 20.27 | 6.17 | 17.68 | 4.59 | 2.59** | 20.96 | 6.45 | 19.09 | 5.58 | 1.87** |
| Dry fodder fed (kg/animal/ day) | 12.32 | 3.76 | 12.85 | 5.14 | -0.53 | 10.71 | 4.11 | 12.03 | 4.90 | -1.32 | 11.52 | 4.50 | 12.44 | 4.56 | -0.93 |
| Concentrate fed (kg/animal/ day) | 3.90 | 1.52 | 3.47 | 1.36 | 0.42 | 3.56 | 1.64 | 3.59 | 1.54 | -0.03 | 3.73 | 1.45 | 3.53 | 1.59 | 0.20 |
| Mineral mixture fed (kg/animal/ day) | 0.03 | 0.04 | 0.02 | 0.04 | 0.00 | 0.02 | 0.03 | 0.03 | 0.04 | -0.01 | 0.03 | 0.04 | 0.03 | 0.04 | 0.00 |
| Average variable cost (₹/animal/ day) | 195.32 | 40.79 | 188.20 | 45.22 | 7.12 | 178.54 | 45.65 | 190.20 | 56.29 | -11.66 | 186.93 | 43.89 | 189.20 | 50.81 | -2.27 |
| Milk price received (₹/kg) | 46.60 | 4.10 | 46.60 | 4.34 | 0.00 | 30.56 | 1.20 | 31.32 | 2.14 | -0.76** | 38.58 | 8.60 | 38.96 | 8.40 | -0.38 |
| Value of milk (000 ₹/animal/lactation) | 110.79 | 22.32 | 111.96 | 25.87 | -1.17 | 95.30 | 33.06 | 102.92 | 41.48 | -7.62 | 103.05 | 29.12 | 107.44 | 34..69 | -4.39 |
| Health score[#] | 3.60 | 0.67 | 3.48 | 0.86 | 0.12 | 3.74 | 0.69 | 3.80 | 0.53 | -0.06 | 3.67 | 0.77 | 3.64 | 0.62 | 0.03 |

**Note:** ** and * indicates statistical significance at 5% and 10%, respectively.

+ 1- Illiterate, 2-Primary (1-5), 3-Middle (6-8), 4-Secondary (9-10), 5-Higher secondary (11-12), 6-Diploma / certification, 7-Graduate and above.

# Health score is an index computed by adding 4 dichotomous variables – artificial insemination (1/0), vaccination (1/0), deworming (1/0) and others (1/0).

**Table 2 Balance test: Linear Probability Estimates**

| Dependent variable: Treatment status (1/0) | (1) Buffalo | (2) Cow | (3) Overall |
|---|---|---|---|
| Incidence of MF (1/0) | -0.131 | -0.008 | -0.064 |
|  | (0.164) | (0.130) | (0.096) |
| Average daily milk yield (kg/animal) | 0.008 | -0.132 | 0.002 |
|  | (0.099) | (0.077) | (0.020) |
| Net income (000' ₹/animal/lactation) | 0.000 | 0.000* | 0.000 |
|  | (0.000) | (0.000) | (0.000) |
| Years of education of dairy farmer | 0.008 | -0.032 | -0.021 |
|  | (0.048) | (0.040) | (0.029) |
| Years of experience in dairying | 0.003 | 0.000 | 0.001 |
|  | (0.007) | (0.006) | (0.004) |
| Land holding (acres) | -0.001 | 0.019 | 0.008 |
|  | (0.010) | (0.013) | (0.008) |
| Herd size (nos.) | 0.014 | 0.008 | 0.008 |
|  | (0.023) | (0.021) | (0.014) |
| Variable costs incurred (₹/animal/day) | -0.001 | 0.004* | 0.000 |
|  | (0.002) | (0.002) | (0.001) |
| Training in dairying (1/0) | 0.006 | 0.089 | 0.061 |
|  | (0.143) | (0.149) | (0.095) |
| Health index of the animal | -0.043 | 0.060 | 0.021 |
|  | (0.082) | (0.093) | (0.062) |
| Extension (contact) index | 0.009 | -0.117 | -0.057 |
|  | (0.070) | (0.071) | (0.049) |
| Constant term | 0.604 | 0.458 | 0.427 |
|  | (0.503) | (0.394) | (0.301) |
| N | 100 | 100 | 200 |
| $R^2$ | 0.03 | 0.11 | 0.03 |
| Joint test of orthogonality (F test) | 0.26 | 1.34 | 0.49 |
| p-value | 0.99 | 0.22 | 0.91 |

**Note:** * $p < 0.1$. Standard errors in parenthesis.

*Attrition*

At the follow-up, a total of 5 animals (2 cows and 3 buffaloes) were sold by the farmers. Therefore, an attrition rate of 2.5% (5 out of 200) was found. These animals were replaced from the stock of 14 animals selected during the baseline survey. The balance was not affected by the replacement of animals.

**EMPIRICAL FRAMEWORK**

*Estimation strategy*

Linear probability model is used to estimate the effect of AMM on MF incidence. Y in Equation 1 is the binary variable with value 1 for MF incidence in follow-up and 0 otherwise. Control covariates ($X_i$) used in the model include baseline MF incidence (1/0), daily milk yield (kg/animal) and parity of animals. $\gamma$ is the parameter of interest as it is the predicted probability of MF occurrence in the treated group in the follow-up.

$$Y_i = \alpha + \gamma Treated + \beta_i . X_i + \varepsilon_i \qquad \ldots (1)$$

We use the Difference in Difference (DID) method to measure the effect of AMM on milk yield and farmer's income. We will compare the outcomes in pre-treatment (baseline) with post-treatment period (follow-up) for animals/households that was in the treatment group with animals/households in the control group.

In our main specification, we estimate the following equation:

$$y_i = \beta_0 + \beta_1 \times period_i + \beta_2 \times treated_i + \beta_3 \times period_i \times treated_i + e_i \qquad \ldots (2)$$

$$y_i = \beta_0 + \beta_1 \times period_i + \beta_2 \times treated_i + \beta_3 \times period_i \times treated_i + \beta_4 \times X_i + e_i \ldots (3)$$

Our coefficient of interest is the interaction term between period (the post-treatment period dummy which takes the value of 1 for observations in follow-up in the main specification) and treated, an indicator of whether the households got the intervention in the post-treatment period. While $\beta_1$ measures the average change in outcome (milk yield/income) between the two time periods for both the treatment and control groups, $\beta_2$ accounts for the permanent difference in outcome between the two (treated and control) groups and can be interpreted as a treatment group specific effect. The coefficient of the interaction term, $β_3$, gives us our DID estimator or the treatment effect, i.e., the causal effect of AMM for the treated group, on outcome between baseline and follow-up, comparative to the control group. $X_i$ is a vector of farmer and herd-level controls. These equations were estimated using STATA routine 'diff' (Villa 2016).

*Identification Strategy*

By the virtue of random assignment and the confirmation from the balance tests, the results are causally identified. Therefore, we estimate the causal effect of AMM (treatment) on milk fever incidence, milk productivity and farmer's income.

**RESULTS AND DISCUSSION**

*Impact of AMM*

Table 3 reports the impact of AMM on incidence of MF. As expected, the effect is negative and larger in cows. Overall, the probability of MF occurrence is 13-15 percentage points lower in treated animals than the control animals. This implies that, if AMM is supplemented, the incidence of MF will reduce to 6-8% from 21% (incidence at baseline = 21%). The reduction in MF will be higher in cows than the buffaloes. At the follow-up, in absolute terms, the incidence of MF reduced from 14% in control buffaloes and 22% in control cows to 2% in treated buffaloes and cows (Figure 1).

**Table 3 Impact of AMM on incidence of MF**

| Strata | Cow | | Buffalo | | Overall | |
|---|---|---|---|---|---|---|
| **Dependent variable** in | Marginal effects (dy/dx) | | | | | |
| **Estimator** | LPM | Probit | LPM | Probit | LPM | Probit |
| **Dependent variable**: | -0.19*** | -0.19 | -0.09* | -0.08 | -0.15*** | -0.13* |
| Incidence of MF | (0.05) | (0.12) | (0.05) | (0.09) | (0.04) | (0.07) |
| Incidence of MF in baseline | 0.27 | 0.27 | 0.15 | 0.15 | 0.21 | 0.21 |
| No. of observations | 200 | 200 | 200 | 200 | 400 | 400 |
| Controls? | Yes | Yes | Yes | Yes | Yes | Yes |

Note: MF in proportion to total animals in the group.
Figures in parenthesis are Delta-method standard errors. * $p < 0.10$, *** $p < 0.01$.
Marginal effects are estimated at treated = 1 and period = 1 (i.e., treatment group at the follow-up), thus are average treatment effects on treated (ATT).
Controls used in cow and buffalo strata – milk yield and lactation order; in overall model – milk yield, lactation order and type of animal (cow=1 and buffalo=0).

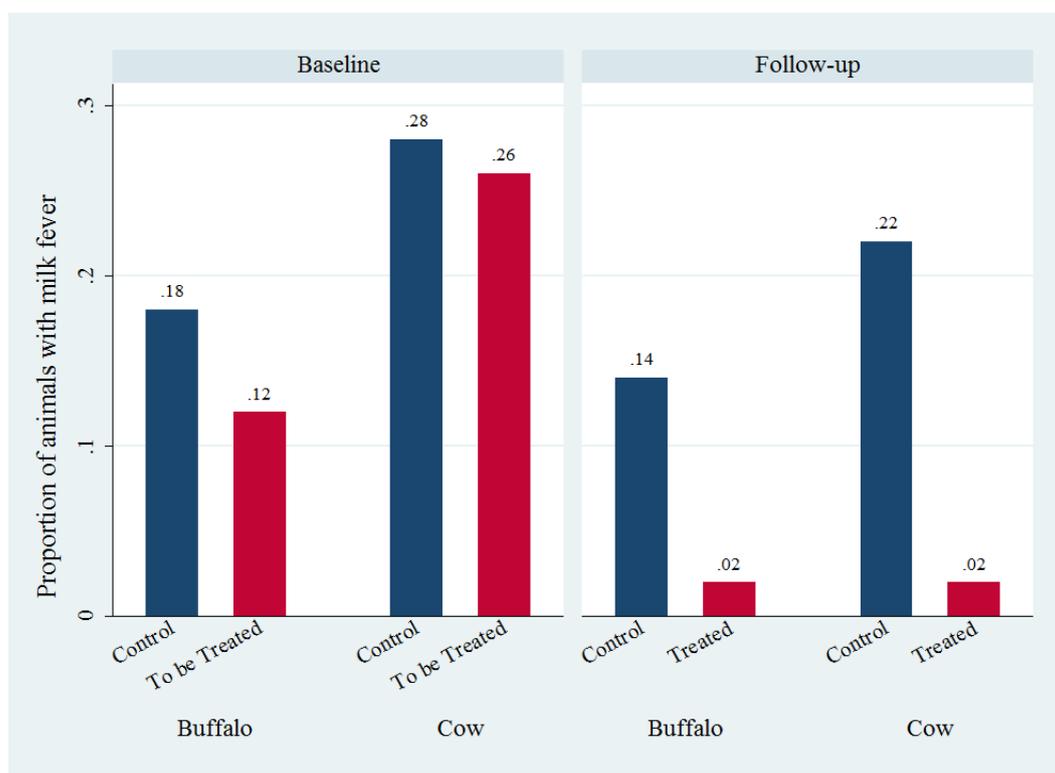

**Figure 1 Impact of AMM on incidence of milk fever**

The impact of AMM on daily average milk yield is presented in Table 4. AMM has a positive effect on milk yield, but the effect is statistically significant in the case of buffaloes and the overall sample. On an average, AMM supplementation increases milk yield by 10.6% and 14.57% in cows and buffaloes, respectively. Overall, the milk yield increases by 11.5% (10.472 to 11.929 kg/animal/day). To check whether the impact estimates are affected by extreme values (a fewer animals driving large differences in milk yield), we plot a kernel density function (Figure 2). Kolmogorov-Smirnov test confirms that the milk yield distribution of treated animals is significantly different than control animals (p < 0.01); implying that the increased milk yield is due to change across the distribution and not just few animals. Thus, the claim of AMM developers (that the milk yield will increase by 10%) is confirmed by our study.

**Table 4 Impact of AMM on milk yield**

| Strata | Cow | | Buffalo | | Overall sample[#] | |
|---|---|---|---|---|---|---|
| Estimator | Difference-in-Differences (DID) | | | | | |
| Dependent variable in | kg/animal/day | Natural logarithm | kg/animal/day | Natural logarithm | kg/animal/day | Natural logarithm |
| Dependent variable: Milk yield | 1.319 (1.100) | 0.106 (0.086) | 1.173*** (0.364) | 0.123*** (0.044) | 1.457** (0.563) | 0.115** (0.048) |
| Mean of Milk yield in baseline | 10.456 | 2.292 | 7.823 | 2.039 | 10.472 | 2.307 |
| No. of observations | 200 | 200 | 200 | 200 | 400 | 400 |
| Controls? | Yes | Yes | Yes | Yes | Yes | Yes |
| $R^2$ | 0.19 | 0.22 | 0.40 | 0.38 | 0.30 | 0.33 |

[#] Dependent variable in overall sample is 3.5% fat corrected milk (FCM). FCM = (0.35 x milk in kg) + (18.57 x fat in kg).
Figures in parenthesis are robust standard errors.
***, ** and * indicates statistical significance at 1%, 5% and 10%, respectively.
DID estimation of milk yield included covariates green fodder fed, dry fodder fed and concentrates fed (all in kg/animal/day) and parity of animals; and of FCM included an additional variable, type of animal (1 – cow, 0 – buffalo).

Our findings of increase in milk yield and decrease in MF incidence corroborate the findings of several experiments done in foreign settings (Lean et al. 2019; Santos et al. 2019; Melendez et al. 2019; Iwaniuk and Erdman 2015; Weich, Block and Litherland 2013). Now that the productivity of Indian dairy animals is on an increasing trend, the risk of MF is also continuously increasing (Appendix 3). This has important implications to the welfare of the farmers; total economic losses (milk production loss + treatment costs + mortality losses) in the sample was estimated at ₹ 4320 (US$ 59) per animal (Cariappa et al. 2021). Therefore, to prevent colossal losses due to MF, the use of AMM is thus recommended by this study.

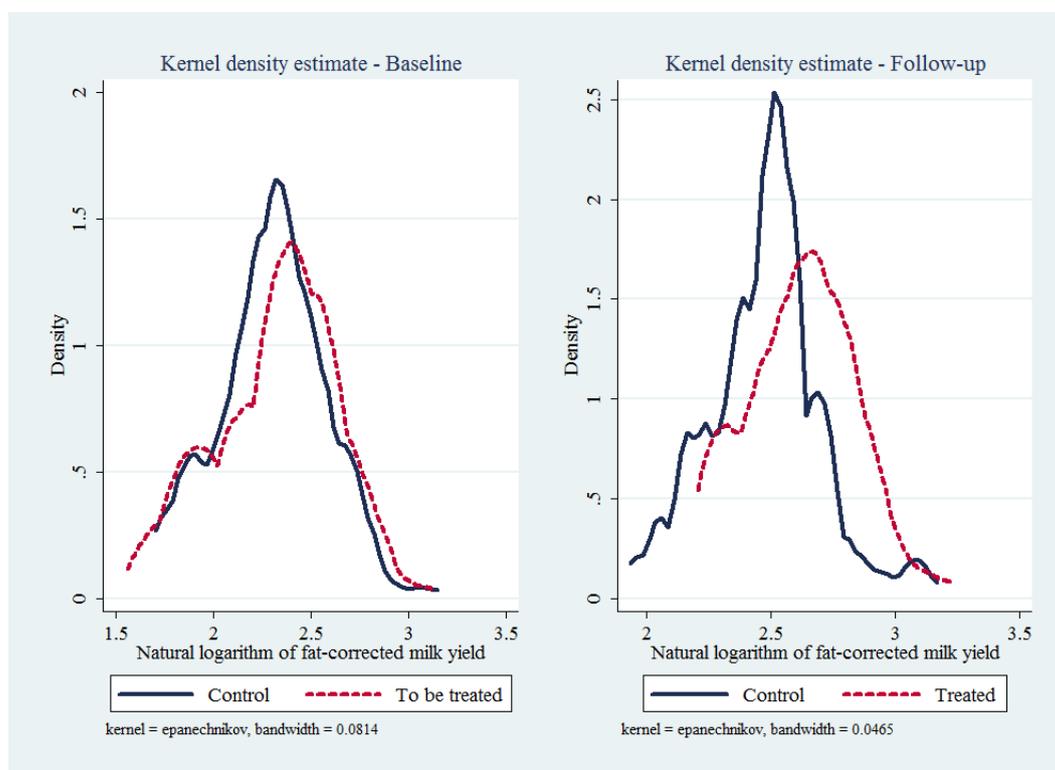

Note: Kolmogorov–Smirnov test statistics showed that the distribution of milk yield for treated and control animals differ at 1% level

**Figure 2 Distribution of 3.5% fat-corrected milk yield in control and treated groups by period**

Impact of AMM on the secondary outcome (farmer's net income) is displayed in Table 5. We find that AMM supplementation leads to a significant increase in the income of the farmer. The impact is higher among buffalo rearers (significant ₹ 17,727 increase per animal per annum) than cow rearers (₹ 14,666). In percentage terms, the increase in returns to variable costs is highest in cow rearers than buffalo rearers with a 52.3% and 24.5% increase compared to the control group, respectively. Overall, farmer's income from dairying increased by ₹ 16,196 per animal per annum, which is a 38% increase compared to the control group. Kolmogorov-Smirnov test for equality of income distributions rejects the null hypothesis of similar distributions of treated and control farmers (Figure 3). This implied that the increase in net income is contributed by all the farmers and not just by some well to do farmers.

**Table 5 Impact of AMM on farmer's net income from dairying**

| Strata | Cow | | Buffalo | | Overall | |
|---|---|---|---|---|---|---|
| Estimator | Difference-in-Differences (DID) | | | | | |
| **Dependent variable** in | ₹/animal/ lactation | Natural logarithm | ₹/animal/ lactation | Natural logarithm | ₹/animal/ lactation | Natural logarithm |
| **Dependent variable: Net income from dairying** | 14666 (9192) | 0.523** (0.262) | 17727*** (5732) | 0.245** (0.119) | 16196*** (5657) | 0.380** (0.149) |
| Mean of income in baseline | 42881 | 10.219 | 52888 | 10.758 | 47885 | 10.496 |
| No. of observations | 200 | 195 | 200 | 200 | 400 | 395 |
| Controls? | Yes | Yes | Yes | Yes | Yes | Yes |
| R-squared | 0.42 | 0.37 | 0.48 | 0.43 | 0.37 | 0.33 |

Note: Net income = gross income – variable cost. Variable costs included are a sum total of expenses on green fodder, dry fodder, concentrates, hired labor including veterinary costs (artificial insemination, vaccination and deworming) and MF treatment costs.
Figures in parenthesis are robust standard errors.
Control covariates – parity and herd size (in nos.), green fodder fed, dry fodder fed and concentrates fed (all in kg/animal/day in linear models and its log values in log transformed models), health score, training on dairying (1/0), experience, land holding, principal income from dairying (1/0) and extension score; and type of animal (1 – cow, 0 – buffalo) in overall equation.
*** and ** indicates statistical significance at 1% and 5%, respectively

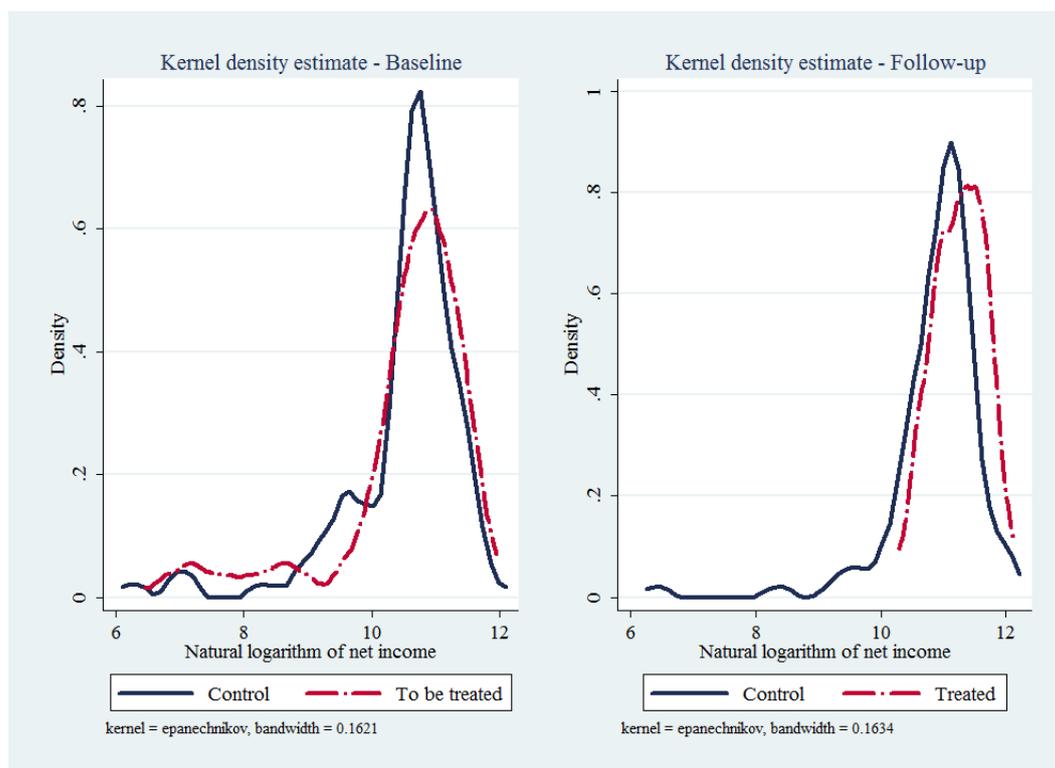

Note: Kolmogorov–Smirnov test statistics showed that the distribution of income for treated and control farmers differ at 1% level

**Figure 3 Distribution of farmer's net income from milk in control and treated groups by period**

*Robustness of the impact estimates*

Results of the robustness test of our impact estimates to alternative specifications are presented in Table 6. DID estimates show that AMM supplementation increased the milk yield by 1.457 kg/animal/day, *i.e.*, an 11.5% increase compared to control animals (Table 3). Table 6, using different specifications (linear-linear, log-linear, linear-log and double log) in OLS estimator, reveals that milk yield increase ranges from 1.50 to 1.68 kg/animal/day and from 11.4% to 12.7%. The range is close to the DID estimates and thus we conclude that our impact estimates are robust to alternative specifications.

**Table 6 Robustness of the impact estimates to alternate specifications**

| Dependent variable: FCM | (1) | (2) | (3) | (4) |
|---|---|---|---|---|
| Estimator | \multicolumn{4}{c}{Ordinary Least Squares (OLS)} | | | |
| Specification | Lin-lin | Log-lin | Lin-log | Log-log |
| **Treated (1/0)** | **1.500*** | **0.114*** | **1.683*** | **0.127*** |
|  | **(0.246)** | **(0.0184)** | **(0.255)** | **(0.0186)** |
| $Y_{t-1}$ (FCM in baseline) | 0.802*** | 0.0572*** |  |  |
|  | (0.0528) | (0.00347) |  |  |
| Green fodder fed (kg/animal/day) | -0.0132 | -0.00197 |  |  |
|  | (0.0212) | (0.00156) |  |  |
| Dry fodder fed (kg/animal/day) | 0.0142 | 0.00138 |  |  |
|  | (0.0317) | (0.00223) |  |  |
| Concentrates fed (kg/animal/day) | 0.120 | 0.00961 |  |  |
|  | (0.0937) | (0.00653) |  |  |
| Parity (nos.) | -0.0246 | -0.00414 | -0.0217 | -0.00376 |
|  | (0.165) | (0.0131) | (0.178) | (0.0137) |
| Animal (1 if cow and 0 if buffalo) | 0.0246 | -0.0392* | 0.338 | -0.0130 |
|  | (0.292) | (0.0204) | (0.353) | (0.0233) |
| Ln $Y_{t-1}$ (ln(FCM) in baseline) |  |  | 8.237*** | 0.602*** |
|  |  |  | (0.680) | (0.0426) |
| Green fodder fed (ln) |  |  | 0.291 | 0.0000772 |
|  |  |  | (0.426) | (0.0313) |
| Dry fodder fed (ln) |  |  | -0.210 | -0.00718 |
|  |  |  | (0.395) | (0.0249) |
| Concentrates fed (IHS) |  |  | 0.622* | 0.0491* |
|  |  |  | (0.307) | (0.0221) |
| Constant term | 3.620*** | 1.904*** | -8.455*** | 1.029*** |
|  | (0.953) | (0.0667) | (2.283) | (0.149) |
| N | 200 | 200 | 200 | 200 |
| $R^2$ | 0.699 | 0.702 | 0.673 | 0.698 |

Note: Standard errors in parentheses. * $p < 0.10$, ** $p < 0.05$, *** $p < 0.01$
The impact of AMM ranged from 12.1% to 12.7% when the dependent variable was changed to absolute milk productivity (not fat-corrected).

*Limitations*

Purposive sampling of animals with high risk of MF (high yielding animals above 2nd parity) might have led the incidence of MF in baseline to be on a higher side and thus leading to an overestimation of impact of AMM. The results are true only for the population similar to our sample and not universal. Therefore, a scale up of this successful pilot or a large cluster level randomized design is required to draw generalizable conclusions.

**SUMMARY AND CONCLUSIONS**

In this study we evaluated the impact of the AMM, a preventive health product against MF in 200 dairy animals. We used a randomized controlled design to estimate the unbiased effects. It was found that the AMM supplementation led to a decrease in MF incidence from 21% in baseline to 2% in treated animals of follow-up. Overall probability of MF occurrence decreased by 13 percentage points in treated compared to control animals. Significant positive effect was found on milk yield and farmer's net income with 12% and 38% increase in treated compared to the control group, respectively.

The benefits of the preventive health product (AMM) are two-fold. First, it decreases the losses and improves animal health (by preventing MF incidence) and second is that it increases animal productivity and farmers' income. To illustrate the implications of these results we do a back of the envelope calculation. The total economic loss due to MF is ₹ 4,320 (US$ 59) per animal (Cariappa et al. 2021). If AMM leads to a decrease in the MF incidence from 21% to 2%, it helps to prevent losses upto ₹ 3,870 (US$ 53). Additionally, AMM leads to a 1.5 kg/day increase in the level of milk production (Table 4). This leads to a significant increase in the farmer's net income to the tune of ₹ 16,196 (US$ 222) per animal per annum (Table 5). Even if the market price of AMM (₹ 540 (US$ 7.4) per animal) is subtracted, net gain stands at ₹ 15,656 (US$ 215). Therefore we conclude that "Prevention of MF (+ ₹ 15,656) is better than cure (- ₹ 4,320)". Further

research building on this successful pilot is required to corroborate the results in different settings to draw generalizable conclusions.

**APPENDIX**

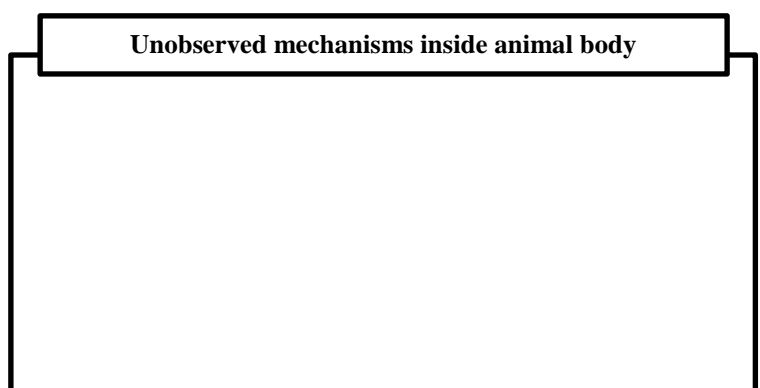

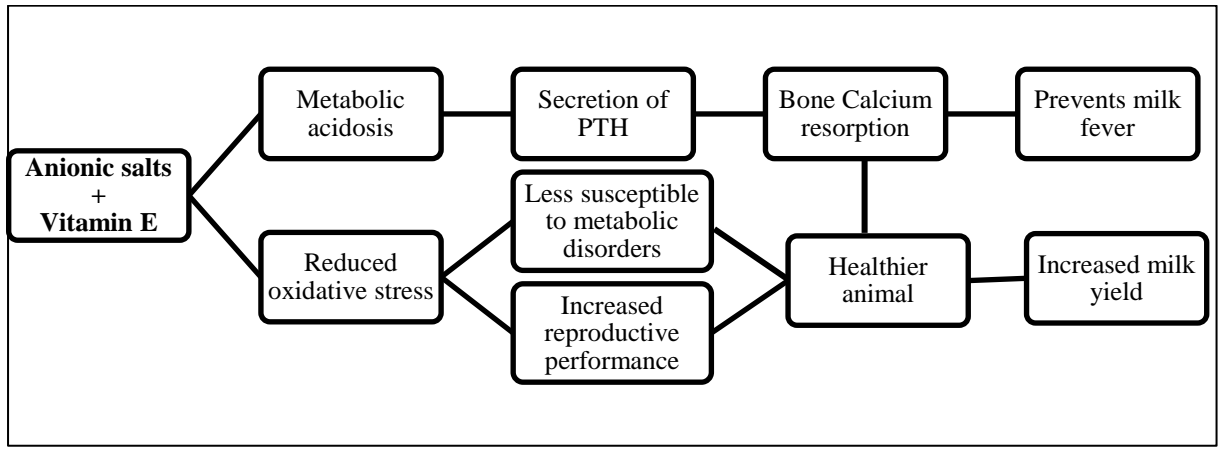

**Appendix 1 How AMM causes MF prevention and increased milk yield**

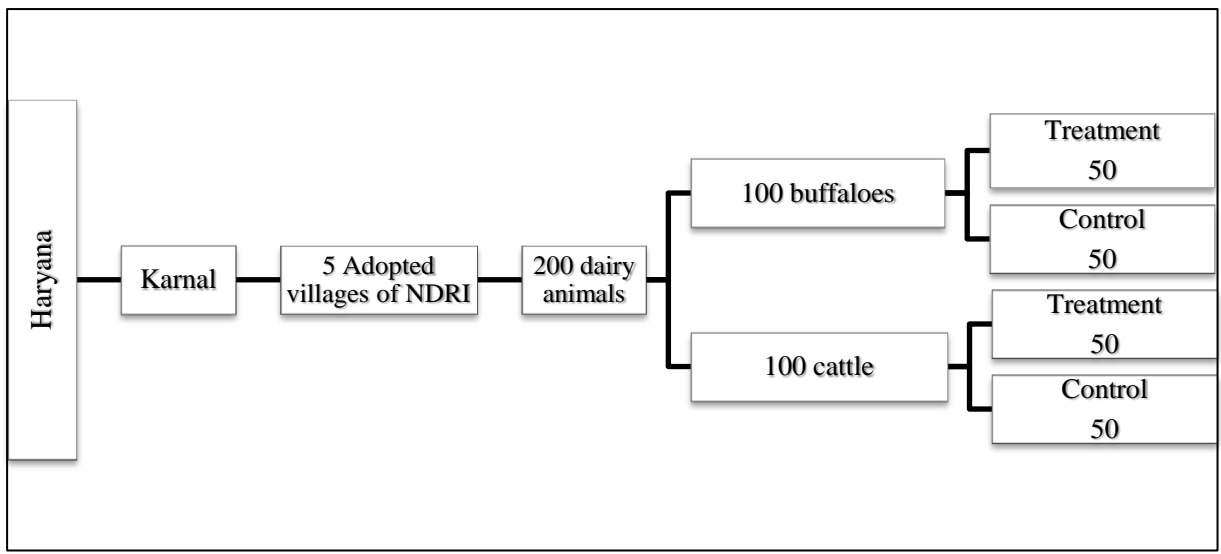

**Appendix 2 Sampling plan of the study**

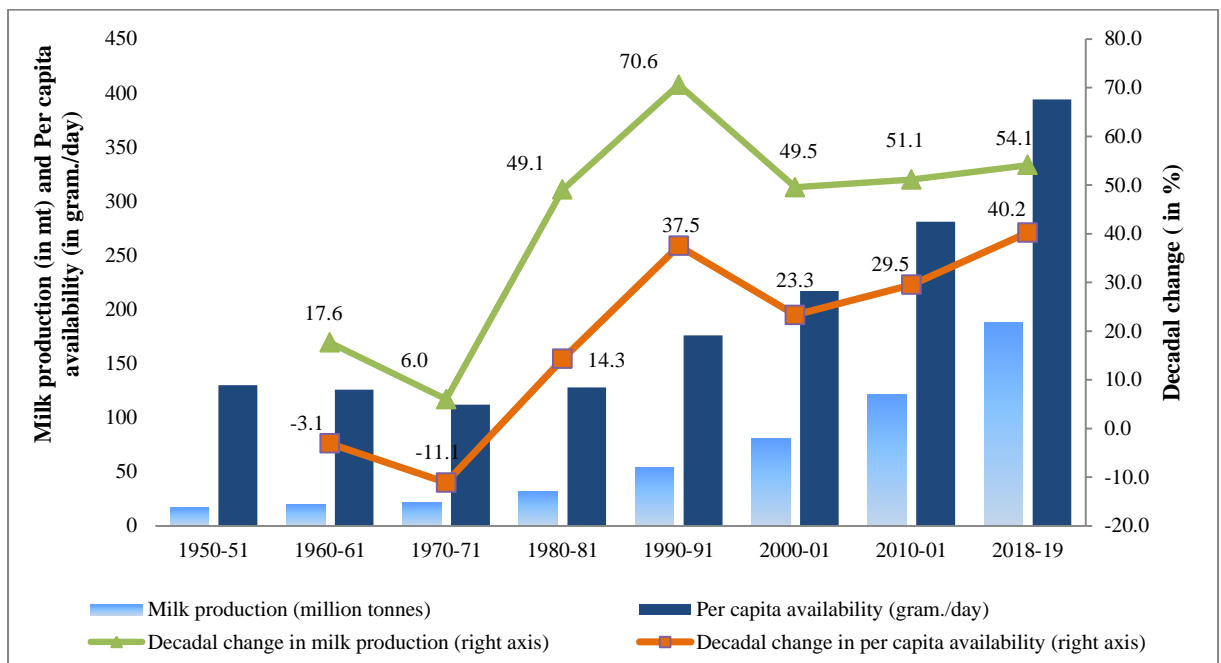

Source: Basic Animal Husbandry Statistics, 2019 (http://dahd.gov.in/sites/default/filess/BAHS%20%28Basic%20Animal%20Husbandry%20Statistics-2019%29_1.pdf)

**Appendix 3 Milk production in India (1950-2018)**